\documentstyle[12pt,epsf]{article}
\def\lsim{\mathrel{\raise.2ex\hbox{$<$}\hskip-.8em\lower.9ex\hbox{$\sim$}}}
\def\gsim{\mathrel{\raise.2ex\hbox{$>$}\hskip-.8em\lower.9ex\hbox{$\sim$}}}
\def\stw{\sin^2\theta_w}
\def\mzs{m_Z^2}
\def\mws{m_W^2}

\def\a{\alpha}
\def\ie{{\rm i.e.}}
\def\O{{\cal O}}
\def\epem{e^+e^-}
\def\R{{\rm R}}
\def\L{{\rm L}}
\def\ksl{k\hskip-6pt/}
\def\qsl{q\hskip-6pt/}
\def\ds{\displaystyle}

\begin{document}

\font\fortssbx=cmssbx10 scaled \magstep2
\hbox to \hsize{
%\special{psfile=uwlogo.ps hscale=8000 vscale=8000 hoffset=-12 voffset=-2}
%\hskip.5in \raise.1in
\hbox{\fortssbx University of Wisconsin - Madison}
\hfill$\vcenter{\hbox{\bf MADPH-96-980}
                \hbox{December 1996}}$ }

\vspace{.5in}

\begin{center}
{\Large\bf Electroweak Interactions: Loops for Cyclists}\,\footnote{Lectures given at the {\it IV Gleb Wataghin School on High Energy Phenomenology}, Universidade Estadual de Campinas, UNICAMP, Brazil, July 1996.}\\[5mm]
F. Halzen\\
\it Physics Department, University of Wisconsin, Madison, WI 53706, USA
\end{center}

\vspace{1in}

\begin{abstract}
We review the ideas of
\begin{itemize}
\item renormalizable field theories,
\item the Standard Model at the Born (neutral currents, the Higgs mechanism and unification) and quantum level.
\end{itemize}
We subsequently illustrate how high statistics experiments are producing the first evidence for the validity of the Standard Model as a spontaneously broken gauge theory.
\end{abstract}

\setcounter{page}{0}
\thispagestyle{empty}

\newpage

\begin{center}
{\Large\bf Electroweak Interactions: Loops for Cyclists}\\[5mm]
F. Halzen
\end{center}

\let\Large=\normalsize

\section{Introduction}

Figure 1 shows a cartoon of the Standard Model, which describes the interactions of quarks and leptons. It is an edifice supported by several pillars. One is extremely solid: gauge theory, which describes particle interactions resulting from symmetry. It is not new, but is built using the framework of renormalizable field theories. Their basic structure was laid out in the 1930's using assumptions that have not been seriously questioned since: quantum mechanics, relativity, and causality. The second pillar is under construction: spontaneous symmetry breaking responsible for the generation of masses of weak bosons. Only two years ago Okun argued that no experimental support existed for the existence of this pillar. We will show that this is no longer true.

\begin{figure}[h]
\centering
\hspace{0in}\epsfxsize=3in\epsffile{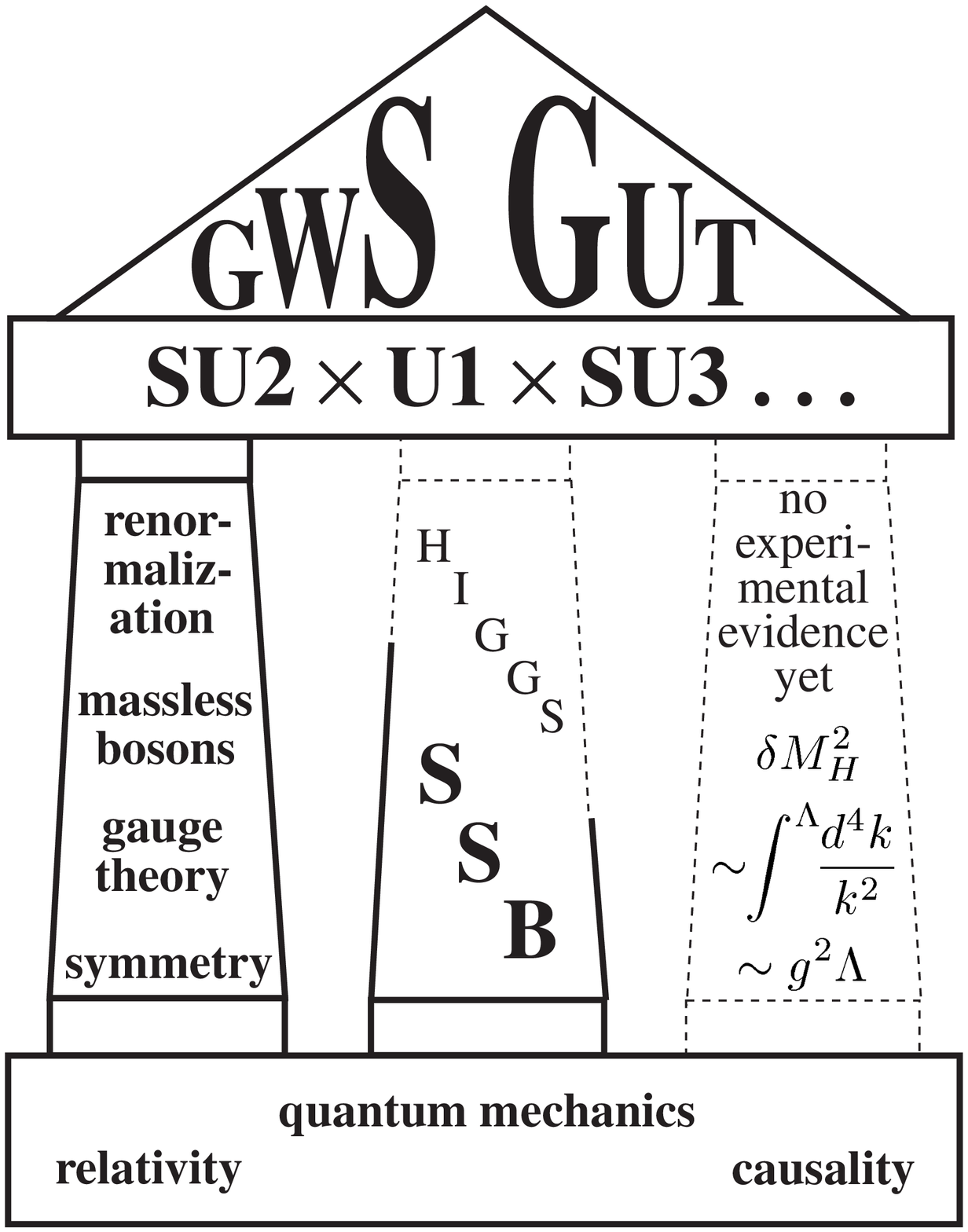}

{\small Figure 1}
\end{figure}

In these lectures we first review the ideas of renormalizable gauge theories. The experimental support for the Standard Model at the Born level is strong and only the main ideas will be discussed: neutral currents, spontaneous symmetry breaking and unification. We will subsequently describe the experimental evidence for the Standard Model as a spontaneously broken field theory. Field theories are verified by measuring radiative corrections. We will illustrate how high statistics experiments are verifying the validity of the quantum structure of the Standard Model.

In the end, we will however conclude that the Standard model, while successfully passing all experimental tests, is incomplete. Pillars are incomplete or missing and we are at present in the unfortunate situation that experiment does not provide us with even a single clue as to what they are made of.

\section{Gauge theory: running couplings}

In gauge theories, symmetries dictate the interactions of quarks and leptons mediated by the exchange of massless vector particles, e.g.\ the photon. Suppose we want to compute the ratio $R$ of the cross sections for the annihilation of electrons and positrons into quarks and muons:
\begin{equation}
R(\alpha,s) = {\sigma(e^+e^-\to q\bar q)\over \sigma(e^+e^-\to\mu^+\mu^-)}\,.
\end{equation}
$R$ is a function of the electromagnetic coupling $\alpha$,
\begin{equation}
\alpha = {e^2\over 4\pi}\,; \qquad \vcenter{\epsffile{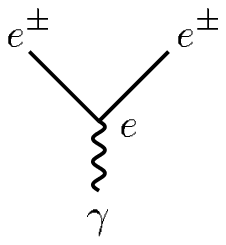}} \label{eq:alpha}
\end{equation}
and the annihilation energy $s= 4E^2_{\rm beam}$:
\begin{equation}
\vcenter{\epsffile{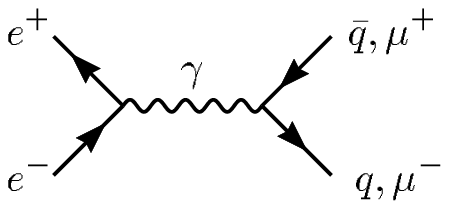}}
\end{equation}
When the annihilation energy far exceeds the light masses $m$ of quarks and leptons, we must expect that for the dimensionless observable $R$,
\begin{equation}
R(\alpha,s) \mathop{\longrightarrow}_{s\gg m^2} \rm constant 
\label{eq:Rconst}
\end{equation}
because there is no intrinsic scale in theories with massless exchange bosons. This prediction disagrees with experiment and is, in fact, not true in renormalizable quantum field theory. The exchange of a massless photon is ultraviolet divergent, requiring the introduction of a cutoff $\Lambda$. Thus a scale is introduced into the calculation and the dimensionless observable $R(\alpha,s,\Lambda^2)$ is of the form
\begin{equation}
R = R\left(\alpha, {s\over\Lambda^2}\right) \,.
\end{equation}
This seems ugly; it is not: $\Lambda$ appears order by order in the perturbative series but not in the final answer. Therefore,
\begin{equation}
\Lambda^2 {dR\over d\Lambda^2} = 0 \,.
\end{equation}
This is the renormalization group equation, which can be written more explicitly:
\begin{equation}
\Lambda^2 {\partial R\over\partial \Lambda^2} + \Lambda^2 {\partial\alpha\over\partial\Lambda^2} {\partial R\over\partial\alpha} = 0 \,,
\label{eq:rge-expl}
\end{equation}
which exhibits that $R$ can depend on $\Lambda$ directly, or via the coupling $\alpha$. Equation~(\ref{eq:rge-expl}) can be rewritten in the variable $t\equiv \ln {s\over\Lambda^2}$. Using $\Lambda^2 {\partial \over \partial\Lambda^2} = -{\partial\over\partial\ln {s\over\Lambda^2}}$\,, we obtain
\begin{equation}
\left( -{\partial\over\partial t} + \beta {\partial\over\partial\alpha} \right)
R\left( \alpha(s),\,{s\over\Lambda^2} \right)= 0 \,,
\label{eq:rge-t}
\end{equation}
where
\begin{equation}
\beta = \Lambda^2 {\partial\alpha\over\partial\Lambda^2} =  {\partial\alpha
\over \partial t} \,. \label{eq:beta}
\end{equation}
With the identification $\Lambda^2=s$, the renormalization group equation has the very simple solution
\begin{equation}
R\Bigl( \alpha(s),\, 1\Bigr) \,.
\label{eq:simple}
\end{equation}
The observable depends on $s$ only via the coupling.
Because $\alpha(s)$ is dimensionless, dimensional analysis requires
\begin{equation}
\alpha(s) = F\left(\alpha(\Lambda^2), {s\over\Lambda^2}\right) \,,
\end{equation}
which is consistent with (\ref{eq:beta}),
\begin{equation}
\Lambda^2 {d\alpha(s)\over d\Lambda^2} = \left[{\partial F\over\partial z} (\alpha(s),z)\right]_{z=1} = \beta(\alpha) \,.
\end{equation}
The solution is
\begin{equation}
t = \ln {s\over\Lambda} = \int_{\alpha(\Lambda)}^{\alpha(s)} {dx\over\beta(x)} \,.
\label{eq:t}
\end{equation}
The ``running" of the coupling is described by the $\beta$-function, which can be computed perturbatively.

In field theory the interaction of 2 electrons by the exchange of a virtual photon is described by a perturbative series
\begin{eqnarray}
&\hidewidth\vcenter{\epsffile{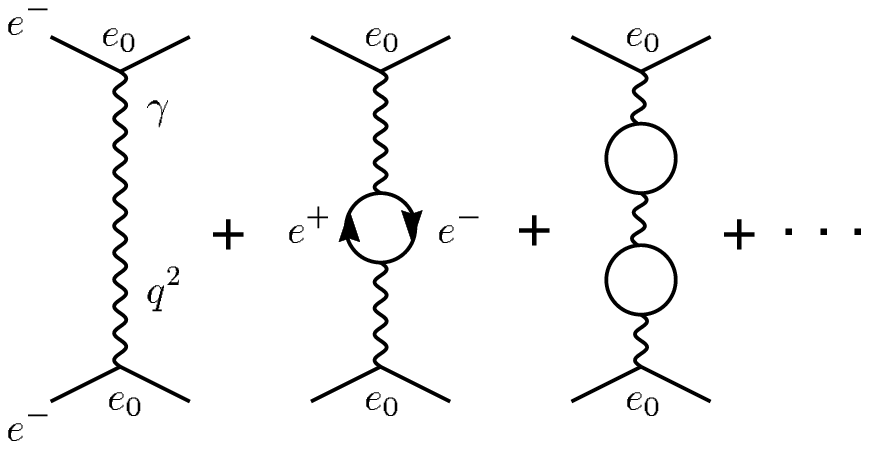}}\hidewidth&\label{eqfig:series}\\
&=& e_0^2 - e_0^2 \pi (q^2) + e_0^2 \pi^2 (q^2) - \cdots, \\
&=& {e_0^2\over 1 + \pi(q^2)  } \,,    \label{eq:summation} \\
\noalign{\hbox{where}}
\pi(q^2) &=& \hskip-1em\vcenter{\epsffile{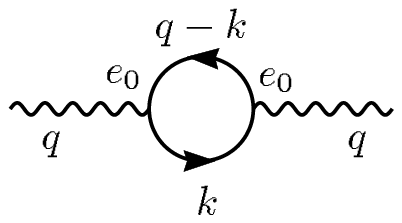}}  \label{eqfig:pi(q^2)}
\end{eqnarray}
Note the negative sign associated with the fermion loop, which is made explicit in order to introduce the summation (\ref{eq:summation}). 

$\pi(q^2)$ is ultraviolet divergent as $k\to\infty$. Explicit calculation\cite{Books} (see Appendix~A) confirms this and we therefore obtain $\pi(q^2)$ in terms of a divergent and finite part
\begin{eqnarray}
\pi(q^2) &=& {e_0^2\over 12\pi^2} \int_{m_e^2}^{\Lambda^2} {dk^2\over k^2} - {e_0^2\over 12\pi^2} \ln {-q^2\over m_e^2} \label{eq:pi(q^2)}
\label{eq:foo}\\
&=& {e_0^2\over 12\pi^2} \;\ln \left(\Lambda^2\over -q^2\right) \,.
\label{eq:QED-result}
\end{eqnarray}
The trick is to introduce a new charge $e$ which is finite:
\begin{eqnarray}
e^2 &=& e_0^2 \left[ 1 - \pi(-q^2=\mu^2) + \cdots \right] \,,
\label{eq:trick-e^2}\\
\noalign{\hbox{or}}
e &=& e_0 \left[ 1 - {1\over2} \pi(-q^2=\mu^2) + \cdots \right] \,.
\label{eq:trick-e}
\end{eqnarray}
We never said what $e_0$ was. It is, in fact, infinitesimal and combines with the divergent loop $\pi$ to yield the finite, physical charge $e$. This operation is performed at some reference momentum $\mu$, e.g.\ $e(\mu=0)$ is the Thomson charge with $\alpha = {e(\mu=0)^2\over4\pi} = (137)^{-1}$. 

To illustrate how this works we calculate $e^-e^-$ scattering. The amplitude is
(ignoring identical particle effects)
\begin{eqnarray}
{\cal M} &=& \vcenter{\epsffile{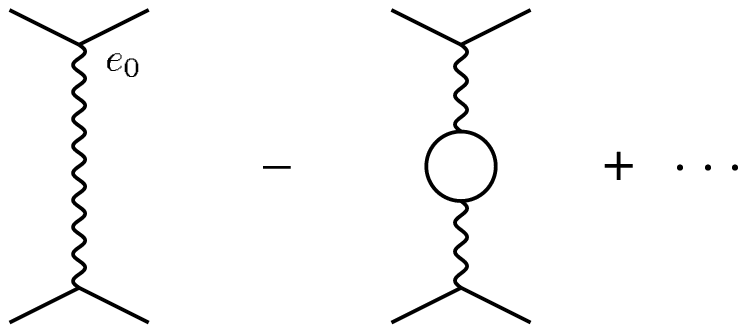}}
\label{eqfig:amp-a}\\
&=& \vcenter{\epsffile{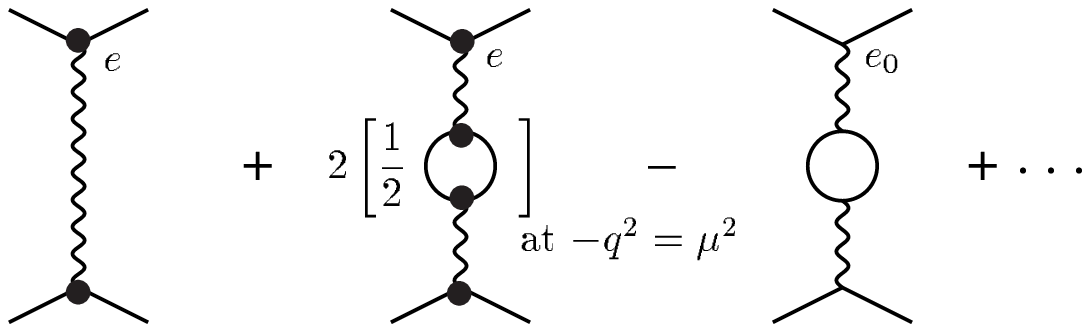}}
\label{eqfig:amp-b}
\end{eqnarray}
where (\ref{eqfig:amp-b}) has been obtained by substituting the renormalized charge $e$ for the bare charge using~(\ref{eq:trick-e}).
\begin{equation}
\vcenter{\epsffile{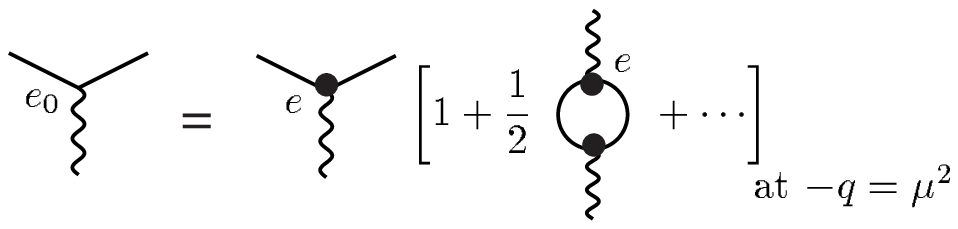}}
\label{eq:sub}
\end{equation}
In the last term of (\ref{eqfig:amp-b}) we can just replace $e_0$ by $e$ as the additional terms associated with the substitution (\ref{eq:sub}) only appear in higher order. Therefore (\ref{eqfig:amp-b}) can be rewritten as:
\begin{eqnarray}
{\cal M} &=& \vcenter{\epsffile{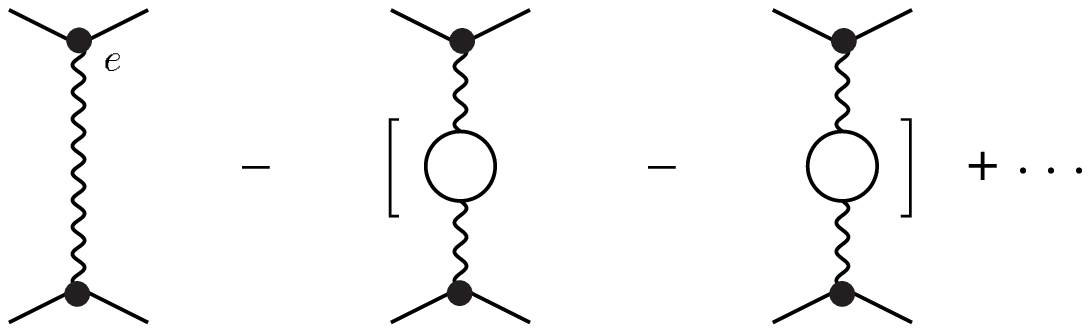}} \label{eqfig:cancel}\\[-5mm]
&& \hskip.9in\epsffile{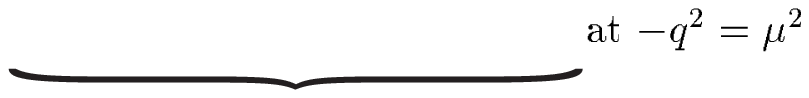}\nonumber\\
&&\hskip1.7in {\alpha\over3\pi} \ln \left(\Lambda^2\over -q^2\right) - {\alpha\over3\pi} \ln \left(\Lambda^2\over-\mu^2\right)\nonumber\\
&&\hskip1.7in ={\alpha\over3\pi} \ln\left(\mu^2\over -q^2\right) = \mbox{finite!} \nonumber
\end{eqnarray}
The divergent parts cancel and we obtain a finite result to ${\cal O}(\alpha^2)$. In a renormalizable theory this cancellation happens at every order of perturbation theory. The price we have paid is the introduction of a parameter $\alpha(\mu^2)$ which is fixed by experiment. The electron charge, unfortunately, cannot be calculated.

In summary,  by using the substitution (\ref{eq:trick-e^2}) the perturbation series using infinitesimal charges $e_0$ and infinite loops $\pi$ has been reshuffled order by order to obtain finite observables. The running charge (\ref{eq:trick-e^2}) can be written as
\begin{equation}
\alpha  = \alpha_0 \left[ 1 - \pi(q^2) + \cdots \right] = {\alpha_0\over 1+\pi(q^2)} \,. \label{eq:alpha-2}
\end{equation}
For the QED result (\ref{eq:QED-result}),
\begin{equation}
\alpha(Q^2=-q^2) = {\alpha_0\over 1-b\alpha_0\,\ln{Q^2\over\Lambda^2}}
\label{eq:alpha-QED}
\end{equation}
with $b=1/3\pi$. The ultraviolet cutoff is eliminated by renormalizing the charge to some measured value at $Q^2=\mu^2$,
\begin{equation}
{1\over\alpha(Q^2)} - {1\over\alpha(\mu^2)} = -b\,\ln{Q^2\over\mu^2} \,.
\label{eq:renorm}
\end{equation}
One also notices that $b$ determines the $\beta$-function to leading order in perturbation theory. We obtain indeed from (\ref{eq:beta}) and (\ref{eq:alpha-QED}) that
\begin{equation}
\beta(\alpha) = {\partial\alpha(Q^2)\over\partial t} = b \alpha^2 + {\cal O}(\alpha^3) \,. \label{eq:beta-2}
\end{equation}

In Table~1 we have listed the $b$-values determining the running of the other Standard Model couplings: the weak couplings $g$, $g'$ and the strong color charge $g_s$. From Eq.~(\ref{eq:simple}) it is clear that much of the structure of the gauge theory is dictated by identifying the momentum dependence of the couplings.

\begin{table}[h]
\def\ds{\displaystyle}
\def\hw{\hidewidth}
\def\q{\quad}
\def\qq{\qquad}
\medskip
\centerline{\small Table 1}\medskip
\centerline{\vrule \vbox{\hsize=5in \hrule  
$$\matrix{
\ds{\rm coupling}\ \ \alpha\equiv{g^2\over4\pi} \vrule depth0pt width0pt height 28pt & \qq\qq b\hbox{-value}\cr
\hrulefill&\qq\qq\hrulefill\cr
\vcenter{\epsffile{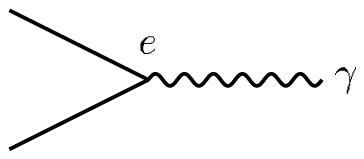}} &\ds \qq\qq{1\over3\pi}\cr
\vcenter{\epsffile{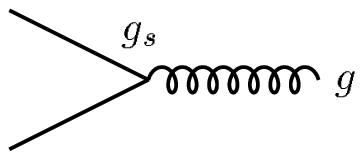}} &\ds \qq\qq{2n_q-33\over12\pi}\cr
\quad\vcenter{\epsffile{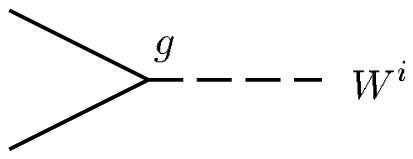}} &\ds \qq\qq{4n_g+{1\over2}n_d-22\over12\pi} \cr
\vcenter{\epsffile{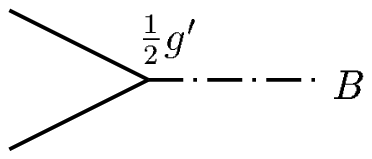}} &\ds \qq\qq{-{20\over3}n_g+{1\over2}n_d\over12\pi}\cr
\multispan2{\hrulefill}\cr
\qquad\qq\q n_q: \hbox{ number of quarks (2--6)}\hw\hfill\cr
\qquad\qq\q n_g: \hbox{ number of generations (3)}\hw\hfill\cr
\qquad\qq\q n_d: \hbox{ number of Higgs doublets (1)}\hw\hfill\cr}$$
\hrule}\vrule}

\end{table}

The formal arguments have revealed the screening of the electric charge. There is physics associated with Eq.~(\ref{eqfig:series}). In quantum field theory a charge is surrounded by virtual $e^+e^-$ pairs which screen the charge more efficiently at large than at small distances. Therefore $\alpha(\mu^2=0)=(137)^{-1}$ is smaller than the short-distance value $\alpha(\mu^2=m_Z^2)=(128)^{-1}$. We note that, qualitatively,
\begin{equation}
{1\over\alpha(0)} - {1\over\alpha(m_Z^2)} \simeq 10 \simeq {1\over3\pi} \,\ln\left(m_Z^2\over m_e^2\right) \,;  \label{eq:1overalpha}
\end{equation}
see (\ref{eq:renorm}).

For 3 generations of quarks the $b$-value for QCD is negative. While $q\bar q$ pairs screen color charge just like $e^+e^-$ pairs screen electric charge (the $2n_q/12\pi$ term in $b$), loops with gluons reverse that effect with a larger, negative $b$-value of $-33/12\pi$. The color charge grows with distance: asymptotic freedom.

\section{Charged and neutral currents}

The electroweak model completes Fermi theory by introducing neutral currents, symmetry breaking and electroweak unification. We briefly review these next.
Fermi theory describes weak interactions via the exchange of a massive weak intermediate boson. The short range of the weak interaction results from of the exchange of a heavy gauge boson of mass $m_W$: 
\begin{eqnarray}
\vcenter{\epsffile{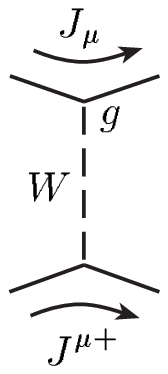}}&=& \left( {g\over\sqrt2} J_\mu \right) {1\over m_W^2} \left({g\over\sqrt2} J_\mu^+ \right) \label{eq:shortrange-a}\\
&=& {4G\over\sqrt2} J_\mu J^{\mu+}\,. \label{eq:shortrange-b}
\end{eqnarray}
The effective coupling $G$ is small compared to the weak coupling $g$ which is of order $e$. The neutral current interaction is described by a coupling $g/\cos\theta_w$,
\begin{eqnarray}
\vcenter{\epsffile{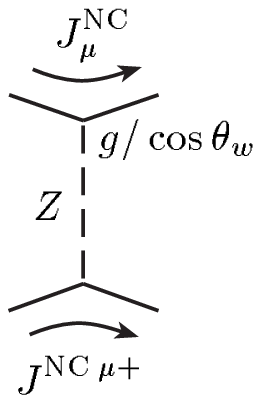}}&=& \left( {g\over\cos\theta_w} J_\mu^{\rm NC}\right) \left(1\over m_Z^2\right) \left({g\over\cos\theta_w} J^{\rm NC \mu+}\right) \label{eq:neutral-a}\\
&=& {4G\over\sqrt2} 2\rho J_\mu^{\rm NC} J^{\rm NC\mu+}\,. \label{eq:neutral-b}
\end{eqnarray}
The relative strength of the neutral and charged currents is parametrized by the weak angle $\cos\theta_w$, or by the $\rho$-parameter as can be seen by comparing (\ref{eq:shortrange-a}) with (\ref{eq:neutral-a}) and (\ref{eq:shortrange-b}) with (\ref{eq:neutral-b}), respectively. Identification of (\ref{eq:shortrange-a}) and (\ref{eq:shortrange-b}) yields
\begin{equation}
{G\over\sqrt 2} = {g^2\over 8m_W^2} \,,  \label{eq:G/sqrt2}
\end{equation}
while combining (\ref{eq:neutral-a}) with (\ref{eq:neutral-b}) gives
\begin{equation}
\rho {G\over\sqrt2} = {g^2\over8m_Z^2\cos^2\theta_w} \,.
\end{equation}
Finally, from the last two equations:
\begin{equation}
\rho = {m_W^2\over m_Z^2 \cos^2 \theta_w} \,.
\label{eq:rho}
\end{equation}
For further discussion it is useful to remember that neutral currents have a coupling $\rho G$ and that $\rho$ represents the relative strength of neutral and charged weak currents, e.g.\ for neutrino-quark scattering:
\begin{equation}
\rho = \quad \vcenter{\epsffile{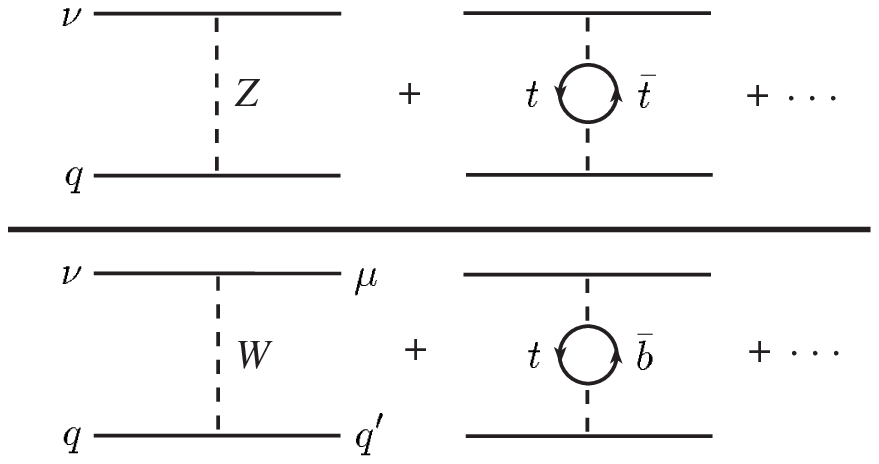}} \label{eqfig:rho}
\end{equation}

\section{Mass of intermediate bosons?}

Describing weak interactions requires the introduction of gauge bosons with mass. Miraculous cancellations like (\ref{eqfig:cancel}) are however a feature of gauge theories with strictly massless exchange particles like the photon. Introducing mass spoils the cancellation and the ultraviolet cutoff $\Lambda$ becomes a parameter which cannot be eliminated. So what? In every order of perturbation theory new parameters $\Lambda',\,\Lambda''\dots$ will appear: a theory with an infinite number of parameters makes no predictions.

The solution is to replace mass by friction. Weak bosons only have an apparent mass which results from the drag they feel from their interaction with a universal medium, like an aether, made of scalar particles. Photons, of course, do not interact with these scalar particles. This solves the problem: while gauge boson mass leads to an unrenormalizable theory, the introduction of such interactions does not. This solution may look a bit contrived but it really is not because scalar particles are required for different, totally unrelated reasons. We will show this in the next section.

A more correct analogy is that the whole universe is filled with a superconducting material. A condensate of scalar particles plays the role of the condensate of Cooper pairs in a real superconductor. The weak gauge fields surrounding quarks and leptons can only penetrate this superconductor over a limited depth, the Meissner effect. This is the origin of the limited range of the weak bosons and their apparent mass.

In order to illustrate the mechanism of mass generation we imagine a world with only photons $(A_\mu)$ and charged scalar particles $(\phi^*,\phi)$. The Lagrangian is
\begin{equation}
{\cal L} = \left[ D^\mu \phi^* D_\mu \phi - \mu^2 \phi^* \phi \right] - \lambda(\phi^*\phi)^2 + \cdots \,,
\end{equation}
with $\lambda>0$ and, as usual, $D_\mu = \partial_\mu - ieA_\mu$. We recognize a theory with scalars of mass $\mu\ (\mu^2>0)$, interacting with a coupling $\lambda$. The Lagrangian represents a different theory when $\mu^2<0$. Writing the Lagrangian in the form
\begin{equation}
{\cal L} = D^\mu \phi^* D_\mu \phi - V(\phi) \,,
\label{eq:lagrangian}
\end{equation}
with
\begin{equation}
V(\phi) = \mu^2 \phi^*\phi + \lambda(\phi^*\phi)^2 \,,
\end{equation}
suggests a theory of massless scalars interacting via the potential $V$. This is not the correct interpretation because of the pedestrian (and incorrect) assumption that the ground state of $V$ is at $\phi=0$, as is usual in old-fashioned textbook examples.

The potential $V(\phi)$ is shown  in Fig.~2 in terms of the real and imaginary components $\phi_1,\phi_2$ of the complex field $\phi$. The minimum of the potential is not at $\phi=0$, but along the circle
\begin{equation}
v^2 = \phi_1^2+\phi_2^2 = {-\mu^2\over\lambda} \,.
\end{equation}
Notice that this indeed requires that $\mu^2<0$. So, in order to interpret the theory correctly we must study the behavior of the field when perturbed along the real and imaginary directions near the minimum $\phi=v$. We do this by rewriting (\ref{eq:lagrangian}) in terms of fields $\eta$ and $\xi$, with
\begin{eqnarray}
\phi(x) &=& {1\over\sqrt2} \left[ v + \eta(x) + i\xi(x) \right] \label{eq:phi}\\
&\simeq& {1\over\sqrt2} (v+\eta) e^{i\xi/v} \,. \label{eq:phi-simeq}
\end{eqnarray}
We obtain
\begin{equation}
{\cal L}' = {1\over2} (\partial_\mu\xi)^2 + {1\over2} (\partial_\mu\eta) - v^2\lambda\eta^2 + {1\over2} e^2v^2 A_\mu A^\mu + \cdots
\label{eq:L'}
\end{equation}
Inspection of $\cal L'$ reveals two real, scalar particles with masses
\begin{eqnarray}
m_\xi &=& 0 \,,\\
m_\eta &=& \sqrt{2\lambda v^2} \,, \label{eq:m-eta}
\end{eqnarray}
and a photon mass term $A_\mu A^\mu$ with
\begin{equation}
m_A = ev \,. \label{eq:photonmass}
\end{equation}
This model illlustrates the essence of the Higgs mechanism where the mass of the gauge particle $A$ is generated by its interactions with scalar particles. The $\eta$ would be called the Higgs boson in the present model.

\begin{figure}[t]
\centering
\hspace{0in}\epsfxsize=3in\epsffile{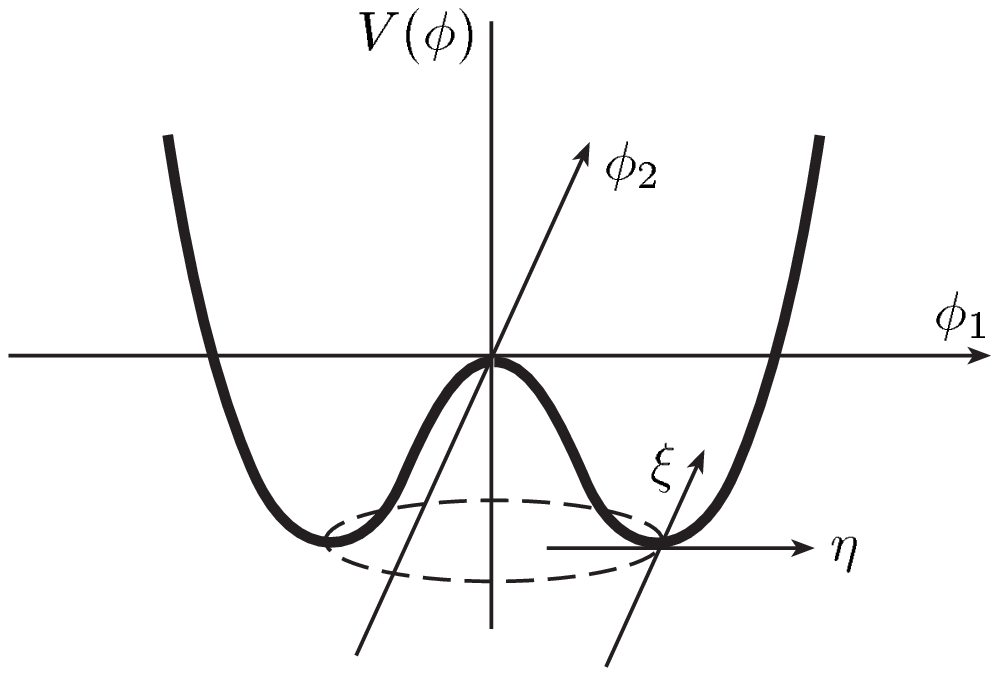}

{\small Figure 2}
\end{figure}

Several comments are appropriate. Note that $\cal L,\,L'$ of Eqs.~(\ref{eq:lagrangian}) and (\ref{eq:L'}) are mathematically identical. If we solved them to all orders of perturbation theory, they would reveal the same physics. Guessing the physics by expansion in the vicinity of the ground state only yields the correct physics interpretation when using the true vacuum (\ref{eq:phi}). No insights can be gained by expanding (\ref{eq:lagrangian}) in the vicinity of $\phi=0$. The scalar particles are frozen into the ground state $\phi=v$ just like Cooper pairs form a condensate below critical temperature. For completeness we should also point out that the massless (Goldstone) particle $\xi$ does, in fact, not appear in the model. A gauge transformation can be used to eliminate it from the Lagrangian as suggested by (\ref{eq:phi-simeq}).

This fantasy world of charged scalars $\phi$ interacting with a gauge vector boson $A_\mu$ illustrates the features of the ``Higgs mechanism": the longitudinal component of $A_\mu$ is generated by interaction with a universal scalar field rather than by the explicit introduction of a mass term. The latter violates gauge invariance and destroys the renormalizability and predictability of the theory. In the Standard Model one introduces a doublet of scalar fields. Each field has 2 components, i.e.\ there are 4 real fields $\phi_1$, $\phi_2$, $\phi_3$ and $\phi_4$ which generate the masses of the 3 gauge bosons $W^+,\ W^-,\ Z$ by the mechanism previously illustrated. Loosely speaking, 3 of the scalar degrees of freedom become the longitudinal components $W_L^+,\ W_L^-,\ Z_L$ of the now ``massive" weak intermediate bosons. The 4th component remains as a physical scalar Higgs particle. It is the elusive witness of the mechanism that generated the weak boson masses; its mass remains undetermined in terms of a parameter $\lambda$; see (\ref{eq:m-eta}). The weak boson masses are given by a generalized version of (\ref{eq:photonmass}):
\begin{eqnarray}
m_W &=& {1\over2} gv \,,\\
m_Z &=& {1\over2} gv \left(1+\tan^2\theta_w\right)^{1/2} \,,\\
v^2 &=& {1\over g^2} 4m_W^2 = \left(\sqrt2 G\right)^{-1} = (246\rm\ GeV)^2\,.
\label{eq:v^2}
\end{eqnarray}
Note that
\begin{equation}
{m_W\over m_Z} = \cos\theta_w \,, \label{eq:costhetaw}
\end{equation}
or, recalling (\ref{eq:rho}),
\begin{equation}
\rho=1 \,.
\end{equation}

Also fermions interact with the scalar particles, thus acquiring masses as well. In renormalizable theories the electron mass, just like the electric charge, remains as an incalculable parameter after renormalization of loops of the form
\begin{equation}
\vcenter{\epsffile{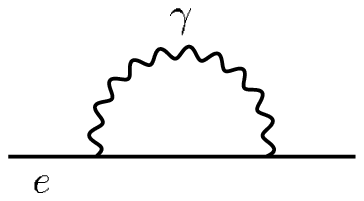}}
\end{equation}
The electron mass is actually determined by its coupling to the scalar field. Masses of all other fermions (including the top with a mass of $m_t\simeq v$) are generated by the Higgs mechanism.

\section{Scalars were already part of the theory!}

One can illustrate this statement simply by calculating the cross section for top quark annihilation into $Z$'s, $t\bar t\to ZZ$, in a Standard Model without scalars. Straightforward Feynmanology yields
\begin{equation}
{d\sigma\over d\Omega}  \left[\; \vcenter{\epsffile{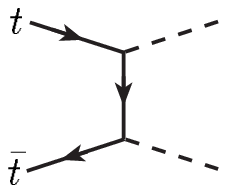}} + \ \vcenter{\hbox{crossed}\hbox{diagram}} \right] ={\alpha^2m_t^2\over m_Z^4} \,.
\label{eqfig:feynmanology}
\end{equation}
We first notice that there is no angular dependence; $d\sigma/d\Omega$ is independent of $\Omega$. The process is purely $S$-wave. We therefore have to conclude that the process violates $S$-wave unitarity, which requires that
\begin{equation}
\sigma_{J=0} \sim {1\over s}\,, \label{eq:unitarity}
\end{equation}
where $s$ is the square of the $t\bar t$ annihilation energy. 

We remind the reader that the unitarity constraint (\ref{eq:unitarity}) simply follows from  the partial wave expansion of the cross section in ordinary quantum mechanics:
\begin{equation}
\sigma = {16\pi\over s} \sum_J (2J+1) \left| f_J \right|^2 \,,
\label{eq:expans}
\end{equation}
with
\begin{equation}
f_J = \exp (i\delta_J) \, \sin\delta_J \,. \label{eq:fJ}
\end{equation}
Here $\delta_J$ are the phase shifts.
Obviously $\left|f_J\right|^2<1$ from (\ref{eq:fJ}) which, when combined with (\ref{eq:expans}), yields
\begin{equation}
\sigma_J < 16\pi (2J+1) {1\over s}
\end{equation}
and (\ref{eq:unitarity}) represents the special case $J=0$. 

The Higgs particle comes to the rescue, introducing the additional diagram:
\begin{equation}
\vcenter{\epsffile{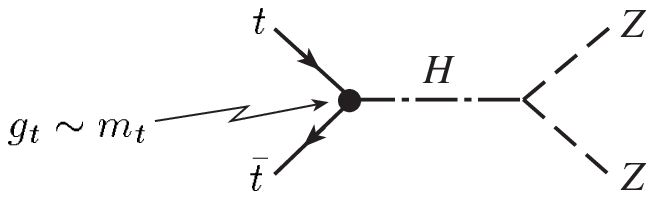}} \sim {g_t^2\over m_Z^4} \,,
\end{equation}
which cancels the ill-behaved $J=0$ term (\ref{eqfig:feynmanology}). The cancellation requires that the top-Higgs coupling $g_t$ (endowing the top quark with mass) satisfies
\begin{equation}
g_t^2 \sim m_t^2 \,,
\end{equation}
a result indeed intrinsic to the Higgs origin of fermion masses. So, if scalars were not invented to solve the problem of mass, they would have to be introduced to salvage unitarity.

We have not found the Higgs particle, but we know that
\begin{equation}
60{\rm\ GeV} < m_H \lsim 1\rm\ TeV\,.  \label{eq:m_H}
\end{equation}
The lower limit can be deduced from unsuccessful searches. Equation (\ref{eq:m-eta}) yields the upper limit
\begin{equation}
m_H = \left(2\lambda v^2\right)^{1/2} < \sqrt2\,v \simeq 350\rm\ GeV\,,
\end{equation}
using (\ref{eq:v^2}). The inequality follows from $\lambda<1$, a requirement which follows from the recognition that the Standard Model's perturbative predictions are correct. This requires couplings to be small, an argument which cannot be taken too literally as it cannot distinguish $\lambda<1$ from $\lambda<4\pi$, for instance. Hence our 1~TeV value quoted in (\ref{eq:m_H}).

\section{Unified electroweak standard model: ``a dirty little model" (J. Bjorken)}

Some 150 years ago Maxwell unified the electric and magnetic forces by postulating the identity of the electric and magnetic charges:
\begin{equation}
\vec F = e\vec E + e_M\, \vec v\times \vec B \,,
\end{equation}
with
\begin{equation}
e = e_M \,. \label{eq:e=e_M}
\end{equation}
Note that the velocity $v$ is the variable which mixes electric and magnetic interactions; when $v\to 0$ magnetic interactions are simply absent but, for charges moving with significant velocity $v$, the two interactions become similar in importance. Unification of the two forces introduces a scale in the mixing variable $v$: the speed of light.

Unification of the electromagnetic and weak interaction follows this pattern with
\begin{equation}
e = g\sin\theta_w\,, \label{eq:e=gstw}
\end{equation}
expressing the equality of electric and weak charge $g$ in terms of the parameter $\theta_w$ introduced in (\ref{eq:neutral-a}). In electroweak theory (\ref{eq:e=gstw}) generalizes (\ref{eq:e=e_M}) to include the weak force. What is the variable mixing electromagnetic and weak forces? At low energy the effects of weak forces between charged particles are swamped by their electromagnetic interaction. At a modern accelerator the weak and electromagnetic forces are equally obvious in the collisions of high energy particles, just like the electric and magnetic forces are in the interaction of high velocity charges. Energy is the mixing variable of electromagnetic and weak forces. The energy scale introduced by their unification is the weak boson mass $m_W$.

The sad reality is that electroweak unification (\ref{eq:e=gstw}) contains a parameter $\theta_w$ which is left to be determined by experiment. The parameter represents the relative strength of charged and neutral currents (cf.\ (\ref{eq:shortrange-a}) and (\ref{eq:neutral-a}) and recall (\ref{eqfig:rho})) as well as the ratio of the weak boson masses $m_W$ and $m_Z$; see (\ref{eq:costhetaw}). The first and only tangible confirmation of electroweak unified theory has been provided by verification that the ratio of the weak boson masses determined at proton-antiproton colliders yields a value of the weak angle which is in agreement with the  value determined in the pioneering neutral current neutrino experiments. On a more mundane level, this common value verifies the doublet nature of the scalar field introduced in the previous section via~(\ref{eq:costhetaw}).

Not until the last two years did true verification of electroweak theory become possible with the first confrontation of its calculated radiative corrections with high statistics measurements performed at the LEP and SLC $e^+e^-$ colliders and at the Fermilab Tevatron. We have barely started down the road of high precision tests familiar from quantum electro-\break dynamics. We describe the first successful steps next.

\section{Quantum structure of the electroweak model: measurements and calculations}

When contemplating the vast amount of evidence for the standard model,
covering strong and electroweak interactions, collider and fixed-target
experiments with lepton, photon, and hadron beams, it is easy to overlook the fact that
verification of the theory at the quantum level is in its infancy, at least by
QED standards. In the electroweak sector familiar tests of the Standard Model
probe the Lagrangian at the Born level: neutrino
neutral and charged current experiments, interference experiments in deep
inelastic scattering, study of $W$  and $Z$ bosons, previously mentioned, and asymmetry measurements in $e^+e^-$ collisions. 

A comprehensive study of the radiative corrections should be a high priority. Implementing such a program can be first formulated from the point of view of the experimentalist. Introducing the notation
\begin{equation}
\stw = s^2=1-c^2\;,\qquad \mws\equiv w\;,\qquad \mzs\equiv z\;,
%%\eqno(3.2)
\end{equation}
electroweak theory predicts at the Born level that:
\begin{eqnarray}
&\ds{\sigma(\nu_\mu e)\over\sigma(\bar\nu_\mu e)} = {3-12s^2+16s^4\over
 1-4s^2+16s^4}\;, \label{eq:nu-ratio}\\  %(3.3)
&\ds{w\over z} = 1-s^2\;, \label{eq:weakmass}\\ %(3.4)
&\ds{\pi\a\over\sqrt{2\,}G_F}{1\over w} = s^2\;, \label{eq:mumass}\\ %(3.5)
&\ds{\Gamma(Z\to f\bar f)\over m_Z} = {\a\over3}\,N_{cf}\left(v_f^2+a_f^2
\right)\;, \label{eq:partialwidth} \\ %(3.6)
&\ds A_{\rm LR} \simeq A_\tau\simeq\left[{4\over3}A_{\rm FB}\right]^{1/2}
\simeq\;2(1-4s^2)\;. \label{eq:ALR} %(3.7)
\end{eqnarray}
Eqs.\ (\ref{eq:nu-ratio})--(\ref{eq:ALR}) represent an incomplete list of experiments capable of measuring $\stw$. Validity of the standard model requires that each measurement yields the same value of $s^2$: (i)~the ratio (\ref{eq:nu-ratio}) of $\nu_\mu$ scattering on left- and
right-handed electrons, which is a function of $\stw$ only, (ii)~the
measurement of the weak boson masses (\ref{eq:weakmass}), previously mentioned, (iii)~the combination of $M_W$,
$\a$, and $G_F$ as determined by the muon lifetime (\ref{eq:mumass}) (we will discuss this relation in detail in the next section), (iv)~the partial
widths (\ref{eq:partialwidth}) of the $Z$ into a fermion pair with vector and axial coupling 
$v_f$ and $a_f$, and color factor $N_{cf}=3\ (1)$ for quarks (leptons), 
and (v)~the various asymmetries (\ref{eq:ALR}) measured at $Z$-factories; see Appendix~B. 

The study of the quantum corrections to the measurements (\ref{eq:nu-ratio})--(\ref{eq:ALR}) is not
straightforward. After inclusion of the $\O(\a)$ corrections, the $\stw$
values obtained from the different methods will no longer be the same because radiative corrections modify (\ref{eq:nu-ratio})--(\ref{eq:ALR}) in different ways.  For example, the diagram
\begin{equation}
\vcenter{\epsffile{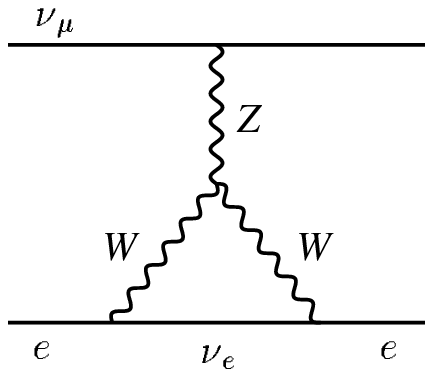}}   %\eqno(3.11)
\end{equation}
modifies the $t$-channel $Z$ propagator measured by (\ref{eq:nu-ratio}); see also~(\ref{eqfig:rho}). It does not, however, contribute to $\O(\a)$ shifts in the $W,\,Z$ masses
\begin{eqnarray}
\vcenter{\epsffile{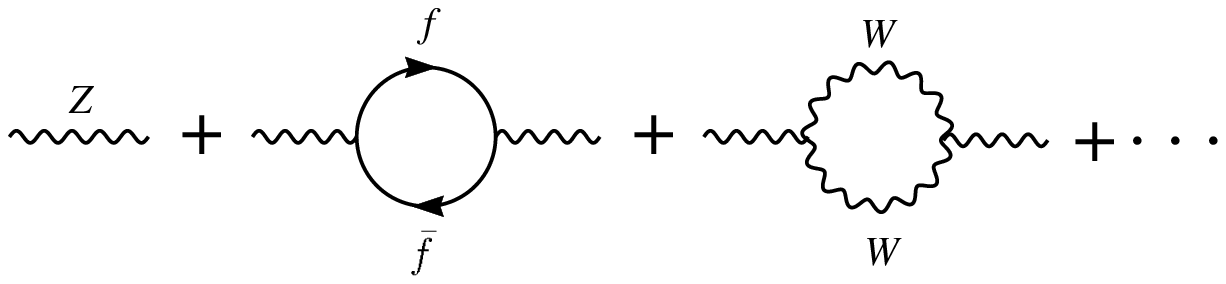}} && \label{eq:Wshift}\\ %(3.12)
\vcenter{\epsffile{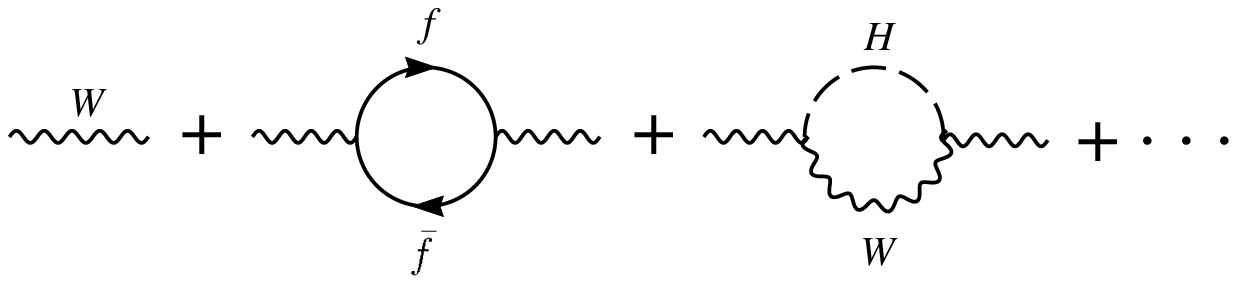}} && \label{eq:Zshift}   %(3.13)
\end{eqnarray}
which yield an improved $\stw$ value via (\ref{eq:weakmass}). There is no real mystery here.
After inclusion of $\O(\a)$  contributions in Eqs.~(\ref{eq:nu-ratio})--(\ref{eq:ALR}), they represent
different definitions of $\stw$. The experimentalist has to make a choice and define the
Weinberg angle to $\O(\a)$ by one of the observables (\ref{eq:nu-ratio})--(\ref{eq:ALR}). Subsequently,
all other experiments should be reformulated in terms of the preferred
``$\sin^2\theta$\rlap.'' What this choice should be is no longer a matter of
debate and we will define $\stw$ in terms of the physical masses of the
weak bosons, \ie
\begin{equation}
\stw\equiv1-{\mws\over\mzs}\;. \label{eq:stw-def} %\eqno(3.14)
\end{equation}
A most straightforward test of the theory is now obtained by fixing (\ref{eq:stw-def})
in terms of the measured weak boson masses and verifying that its
value coincides with the value of $\sin^2\theta_w$ obtained from an analysis of $\nu$ deep-inelastic
scattering data using the $\O(\a)$ prediction for (\ref{eq:nu-ratio}) written in terms of (\ref{eq:weakmass}). The same procedure can be repeated for the other measurements of $\theta_w$, e.g., (\ref{eq:mumass}), (\ref{eq:partialwidth}) and (\ref{eq:ALR}). 

The choice (\ref{eq:stw-def}) is particularly useful in that one can estimate the radiative corrections in terms of the renormalization group, which has been previously introduced. The $\O(\a)$ corrections can be qualitatively understood in terms of the loop corrections to the vector-boson propagators (\ref{eq:Wshift}) and (\ref{eq:Zshift}).
 In a more technical
sense the choice (\ref{eq:stw-def}) is closely related to the use of the on-mass-shell (OMS) renormalization scheme, which generalizes the renormalization techniques, introduced for electrodynamics, in a straightforward way to the electroweak model. 

Renormalization techniques take care of
UV divergences appearing in gauge theories at the quantum level. In the
introduction we illustrated how the divergence in the photon vacuum
polarization is absorbed into the Thomson charge. We pay a price: the Thomson
charge is no longer predicted and the  charge is renormalized to
its measured value at $q^2=0$. Not all predictive power is lost. The screening of the charge $\a(q^2)$ can still be predicted and confronted with experiment. All UV divergences in QED can be absorbed in two parameters, $\a$ and $m_e$. It is eminently reasonable to copy this scheme for calculations in electroweak
theory. The list of parameters, to be fixed by experiment, now includes
\begin{equation}
\a,\; m_W,\; m_Z,\; m_H,\; m_f\,, \label{eq:parameters} %\eqno(3.15)
\end{equation}
where  $m_f$ represents the lepton and  quark
masses $m_e,\ldots, m_t$. The weak mixing angle $\stw$ does not appear in the
list of parameters; its value is automatically determined by $m_W,\,m_Z$ via
(\ref{eq:stw-def}). For some this procedure may seem unfamiliar. Traditionally the Standard Model Lagrangian is determined in terms of 
\begin{equation}
 g,\; g',\; \lambda,\; \mu,\; g_f, \label{eq:L-terms} %\eqno(3.16)
\end{equation}
which represent the bare electroweak couplings, the parameters of the minimal
Higgs potential, and the ``Yukawa" couplings of the Higgs particle to fermions. There is no mystery here. In 
principle any choice will do. There is, in fact, a direct translation between
sets (\ref{eq:parameters}) and (\ref{eq:L-terms})
\begin{equation}
\begin{array}{rcl}
g^2&=&\ds e^2{z\over z-w}\;, \\[4mm]  %(3.17) 
g'^2&=&\ds e^2{z\over w}\;, \\[4mm]  %(3.18)
\lambda&=&\ds e^2{z M_H^2\over8w(z-w)}\;, \\[4mm] %(3.19)
g_f&=&\ds e^2{z m_f^2\over2w(z-w)}\;. %(3.20)
\end{array}
\end{equation}

\section{The lifetime of the muon}

As an example we will show how the relation (\ref{eq:mumass}) is calculated to ${\cal O}(\alpha)$ in terms of the weak angle $\theta_w$ defined by (\ref{eq:weakmass}).  The origin of the relation (\ref{eq:mumass}) is the muon's lifetime which, to leading order, is given by the diagram
\begin{equation}
\Gamma_\mu^{(0)} = \vcenter{\epsffile{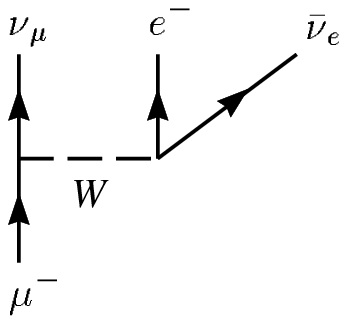}}
\end{equation}
In Fermi theory, electromagnetic radiative corrections must be included to obtain the result to ${\cal O}(\alpha)$. Symbolically,
\begin{equation}
\Gamma_\mu^{(1)} ={G\over\sqrt2}\; [ 1 + \mbox{photonic corrections} ] \,,
\label{eq:Gamma-fermi}
\end{equation}
where
\begin{equation}
\mbox{photonic corrections} = \vcenter{\epsffile{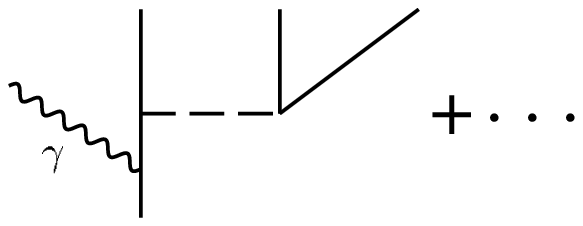}} 
\end{equation}
In electroweak theory, on the other hand,
\begin{equation}
\begin{array}{rcl}
\Gamma_\mu^{(1)} = \displaystyle{e^2\over 8s^2c^2z}\; \Big[\;1 &+&\rm photonic\ corrections\\
& +&\rm propagator\\
& +&\rm vertex\\
& +&\rm box\;\Big]
\end{array} 
\label{eq:Gamma-ew}
\end{equation}
where
\begin{eqnarray}
\rm propagator &=& \vcenter{\epsffile{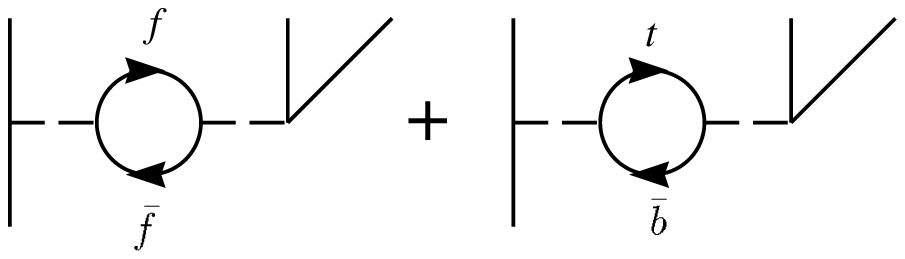}}\nonumber\\
&&\vcenter{\epsffile{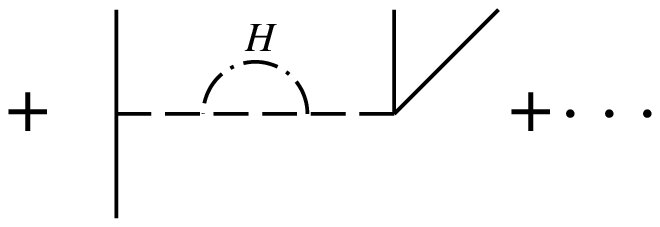}}\\
\rm vertex &=& \vcenter{\epsffile{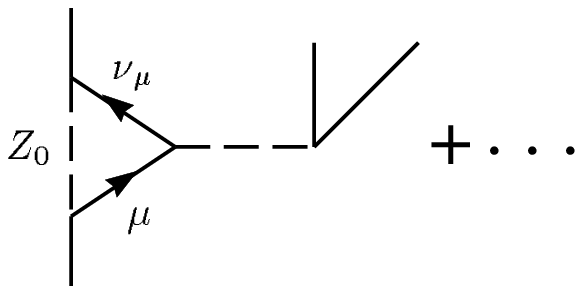}}\\
\noalign{\hbox{and}}
\rm box &=& \vcenter{\epsffile{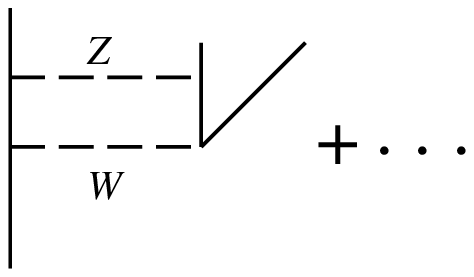}}
\end{eqnarray}
Equating (\ref{eq:Gamma-fermi}) and (\ref{eq:Gamma-ew}) we obtain
\begin{equation}
{G\over\sqrt2} = {e^2\over 8s^2c^2z} (1 + \Delta r) \,, \label{eq:G}
\end{equation}
with
\begin{equation}
\Delta r = \Delta \alpha - {c^2\over s^2} \Delta\rho + \Delta_{\rm rem} \,.
\label{eq:Deltar}
\end{equation}
We note that the purely photonic corrections drop out. The electroweak radiative corrections are gathered in $\Delta r$. Notation (\ref{eq:Deltar}) recognizes the fact that in the OMS scheme, vacuum polarization loops dominate this quantity. We specifically isolated the fermions which are responsible for the running of $\alpha$ from the muon to the $Z$ mass,
\begin{equation}
\Delta\alpha = \sum_f \hskip-.5em\vcenter{\epsffile{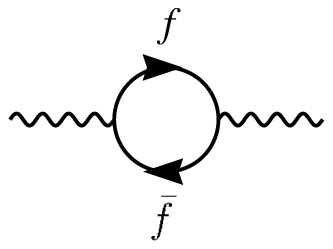}} 
\label{eq:Deltalpha}
\end{equation}
as well as the third generation, heavy quark diagram
\begin{equation}
\Delta\rho = \hskip-.5em\vcenter{\epsffile{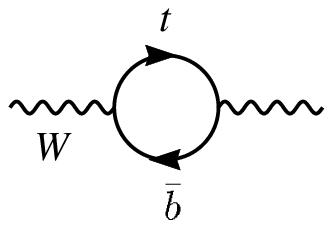}} 
\label{eqfig:Drho}
\end{equation}
Other contributions are small in the OMS scheme and are grouped in the ``remainder" $\Delta_{\rm rem}$.

Before discussing the status of measurements of $\Delta r$, we make several comments. To leading order $\Delta r = 0$ and,  using (\ref{eq:alpha}) and (\ref{eq:costhetaw}), (\ref{eq:G}) reduces to the Born relation (\ref{eq:mumass}). The full order $\alpha$ calculation of $\Delta r$ will not be presented here. It is straightforward and relatively simple. The complete result has been written in analytic form in Ref.~\cite{ParticleWorld}. To the extent that $\Delta_{\rm rem}$ is small, one can imagine summing the series
\begin{equation}
\vcenter{\epsffile{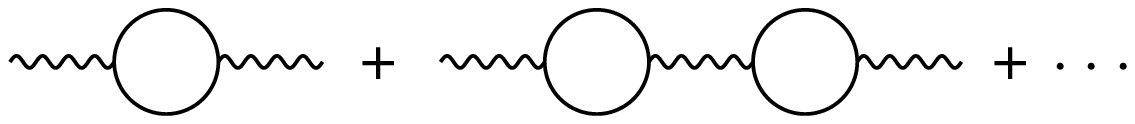}}
\end{equation}
by the replacement $(1+\Delta r) \to (1-\Delta r)^{-1}$ in (\ref{eq:G}).

We already discussed the running of $\alpha$ from the small lepton masses to $m_Z$; see (\ref{eq:1overalpha}). The other large contribution $\Delta\rho$, which  represents the loop (\ref{eqfig:Drho}), is our primary focus here. Its value is given by\cite{Books,ParticleWorld}
\begin{equation}
\Delta\rho = {\alpha\over4\pi} {z\over \omega(z-\omega)} N_C |U_{tb}|^2 \left[ m_t^2 F(m_t^2, m_b^2) + m_b^2 F(m_b^2, m_t^2) \right] \,,
\label{eq:Drho-value}
\end{equation}
with
\begin{equation}
F(m_1^2, m_2^2) = \int_0^1 dx \,x \,\ln\left[ m_1^2(1-x) + m_2^2 x \right] \,.
\label{eq:F}
\end{equation}
Here the number of colors $N_C=3$ and $U_{tb}$ is the KM matrix element; $|U_{tb}|^2\simeq1$. The diagram has the important property that, defining $m_t = m_b+\epsilon$, 
\begin{equation}
\Delta\rho \simeq {G\over 3\pi^2}\;\epsilon \,.
\label{eq:epsto0}
\end{equation}
So in QED, where only equal mass fermions and antifermions appear in neutral photon loops, $\epsilon=0$ and diagrams of this type are not possible. They are, in fact, prohibited in QED by general arguments. This can be seen by rewriting  (\ref{eq:Drho-value}) and (\ref{eq:F}) in the form
\begin{eqnarray}
\Delta\rho &=& {G\over4\pi} \left[ m_t^2 + m_b^2 - {2m_b^2m_t^2\over m_t^2-m_b^2} \ln{m_t^2\over m_b^2} \right] \nonumber\\
&\simeq& {G\over4\pi} m_t^2 \simeq {3\alpha\over16\pi} {1\over c^2s^2} {m_t^2\over z} \,.  \label{eq:SMmt}
\end{eqnarray}
The appearance of a $m_t^2/z$ contribution to an observable is far from routine. It is indeed forbidden in QED and QCD where virtual particle effects are suppressed by ``inverse" powers of their masses; (\ref{eq:epsto0}) embodies this requirement because $\epsilon=0$ for photon loops. Conversely, the appearance of an $m_t^2/z$ term is a characteristic feature of the electroweak theory. $\Delta\rho$ provides us with a most fundamental probe of electroweak theory short of discovering the Higgs boson. 

We are now ready to illustrate that $\Delta\rho\neq0$ and is, in fact, consistent with the Standard Model value (\ref{eq:SMmt}) calculated using the experimental value of the mass of the recently discovered top quark. We first determine the experimental value of $\Delta r$ from (\ref{eq:G}). Using (\ref{eq:G/sqrt2}) and (\ref{eq:e=gstw}):
\begin{equation}
\Delta r_{\rm exp} = 1 - (37.281)^2 {z\over \omega(z-\omega)} = 0.037 \,.
\label{eq:Dr-exp}
\end{equation}
Here, and in what follows, we perform calculations with
\begin{equation}\def\arraystretch{1.2}
\begin{array}{rcl}
\alpha^{-1} &=& 4\pi/e^2 = 137.0359895 \,,\\
m_Z &=& 91.1885 \pm 0.0022 \,,\\
m_W &=& 80.346 \pm 0.046^{+0.0012}_{-0.0021} \,,\\
\sin^2\theta_w &=& 0.2237\pm 0.0009^{+0.0004}_{-0.0002} \,,\\
m_t &=& 178\pm 8 ^{+17}_{-20} \,.
\end{array}\label{eq:values}
\end{equation}
We next recall (\ref{eq:1overalpha}):
\begin{equation}
\Delta\alpha \simeq 1 - {\alpha(0)\over\alpha(m_Z^2)}
= 1 - {128\over137} = 0.071 \,.
\label{eq:Da}
\end{equation}
The crucial point is that $\Delta r_{\rm exp} \neq \Delta\alpha$; cf.~(\ref{eq:Dr-exp}) and (\ref{eq:Da}). The ${\cal O}(\alpha)$ Standard Model relation (\ref{eq:Deltar}) requires a non-vanishing value of $\Delta\rho$. Using (\ref{eq:SMmt}) and (\ref{eq:values}), we obtain that $\Delta\rho=0.0094$ and (\ref{eq:Deltar}) yields
\begin{equation}
(\Delta r)_{\rm calculated} = \Delta\alpha - {c^2\over s^2}\Delta\rho = 0.0384 \,, \label{eq:Dr-calc}
\end{equation}
in agreement with the experimental value (\ref{eq:Dr-exp}). We leave it as an exercise to insert errors into the calculation and show that our argument survives a straightforward statistical analysis. 

As recently as 2 years ago Okun could, correctly, make the claim that the quantum structure of the electroweak model was untested beyond the well-known running of the couplings. This is no longer true as experimental errors are such that $\Delta r < \Delta\alpha$, with the difference accounted for by a value of $\Delta\rho$ which is consistent with the Standard Model prediction. We checked that an essentially electroweak radiative correction is predicted in agreement with experiment.

The Higgs particle makes a contribution to $\Delta r$:
\begin{equation}
\Delta h = \vcenter{\epsffile{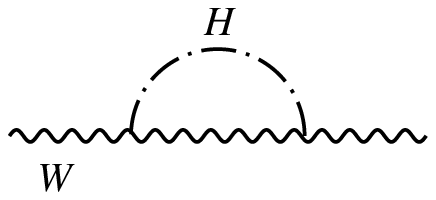}} = {11\alpha\over48\pi} {1\over c^2} \, \ln{m_H^2\over z} \,.
\label{eqfig:Dh}
\end{equation}
From (\ref{eq:m_H}) we obtain that $\Delta h < 0.0006$, a contribution too small to be sensed by the simple analysis presented above. The quantity $\Delta r$ is in principle sensitive to the Higgs mass. More sophisticated analyses which include the dominant ${\cal O}(\alpha^2)$ corrections are now yielding weak, but definite, constraints on the value of $m_H$.

Other measurements support the electroweak model's radiative correction associated with the $t\bar b$ loop $\Delta\rho$. Recall that charged weak currents couple with strength $G$, while neutral currents couple as $\rho G$; see (\ref{eq:shortrange-b}) and (\ref{eq:neutral-b}). The neutral current decay of $Z$ into neutrinos is therefore proportional to $\rho G$:
\begin{equation}
\Gamma(Z\to \nu\bar\nu) = (\rho G) {3\sqrt2\over24\pi} m_Z^3 \,.
\end{equation}
The measured value of $499.9\pm2.5$~MeV is larger than the value calculated from the above equation which is 497.9, although the statistics are not overwhelming. Nevertheless, the loop contribution (\ref{eq:SMmt}) increases $\rho$ to a value $1+\Delta\rho=1.0094$, bridging the gap. In the end a professional approach follows the technique we previously mentioned: generalize the theoretical expressions for the observables (\ref{eq:nu-ratio})--(\ref{eq:ALR}) to 1-loop and show that all measurements yield a common value of $\sin^2\theta_w$. The impressive results of such an analysis are shown in the following table. We have attempted to describe the full formalism in a relatively accessible way in Ref.~\cite{Brazil}.

\begin{table}[h]
\centering
\parbox{5.5in}{\small Table 2: Values obtained for $s_W^2$ (on-shell) and $\widehat s_W^2$ ($\overline{\rm MS}$) from various reactions assuming the global best-fit value $m_t=180\pm07$~GeV (for $M_H=300$~GeV), and $\alpha_s=0.123\pm0.004$. The uncertainties include the effect of 60~GeV${}<M_H<1$~TeV. The determination from $\Lambda_Z,\ R$, and $\sigma_{\rm had}$ uses the experimental value of $M_Z$, so the values obtained are from the vertices and not the overall scale. Table taken from the 1996 Particle Data Book.}
\medskip
\hspace{0in}\epsffile{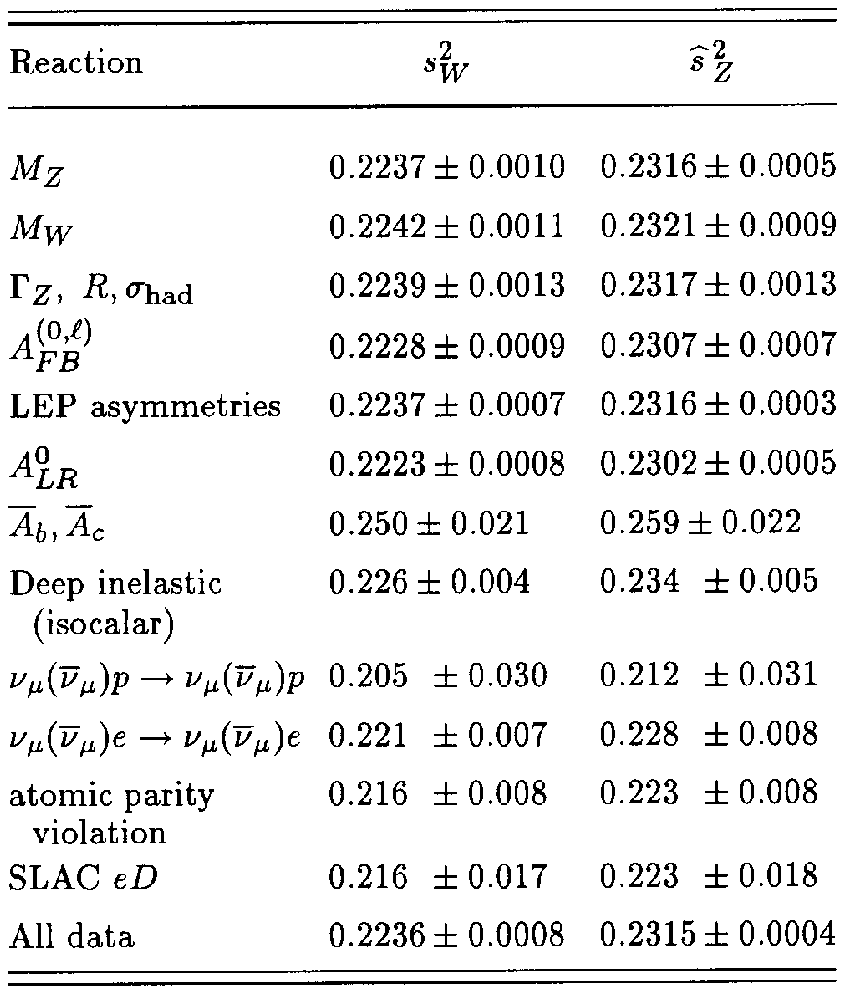}
\end{table}

\section{Outlook}

In the last 2 years the radiative corrections predicted by the Standard Model have successfully confronted experiment. Vacuum polarization effects associated with $t\bar b$ loops are a new and characteristic feature of electroweak theory and their contribution has been established experimentally. Virtual effects proportional to the square of the fermion mass do not exist in QED. Their observation (where the unified electroweak model is concerned) is reminiscent of, and similar in importance to, the first observation of virtual effects associated with $e^+e^-$ pairs via the Lamb shift. The program is however far from complete. It will not have escaped the reader's attention that the precision of the confrontation between theory and experiment is limited by the relatively large errors on the measurements of $m_W$ and $m_t$. The problem can be quantified by rewriting (\ref{eq:Dr-exp}) and (\ref{eq:Deltar}) as
\begin{equation}
\Delta r_{\rm exp} = F(m_W, m_t, m_H) \,, \label{eq:F(m's)}
\end{equation}
using (\ref{eq:SMmt}), (\ref{eq:Da}) and (\ref{eqfig:Dh}). This relation between $m_W$ and $m_{\rm top}$ is shown in Fig.~3  for a variety of choices for $m_H$. Also shown are the latest Fermilab Tevatron results. The hope is to perform measurements of $m_W, m_t$ to a precision $\Delta m_W<50$~MeV and $\Delta m_t < 5$~GeV. Such a measurement should determine $m_H$. Confirmation of the Standard Model will require the detection of the Higgs particle at that value. 

\begin{figure}[h]
\centering
\hspace{0in}\epsfxsize=3.7in\epsffile{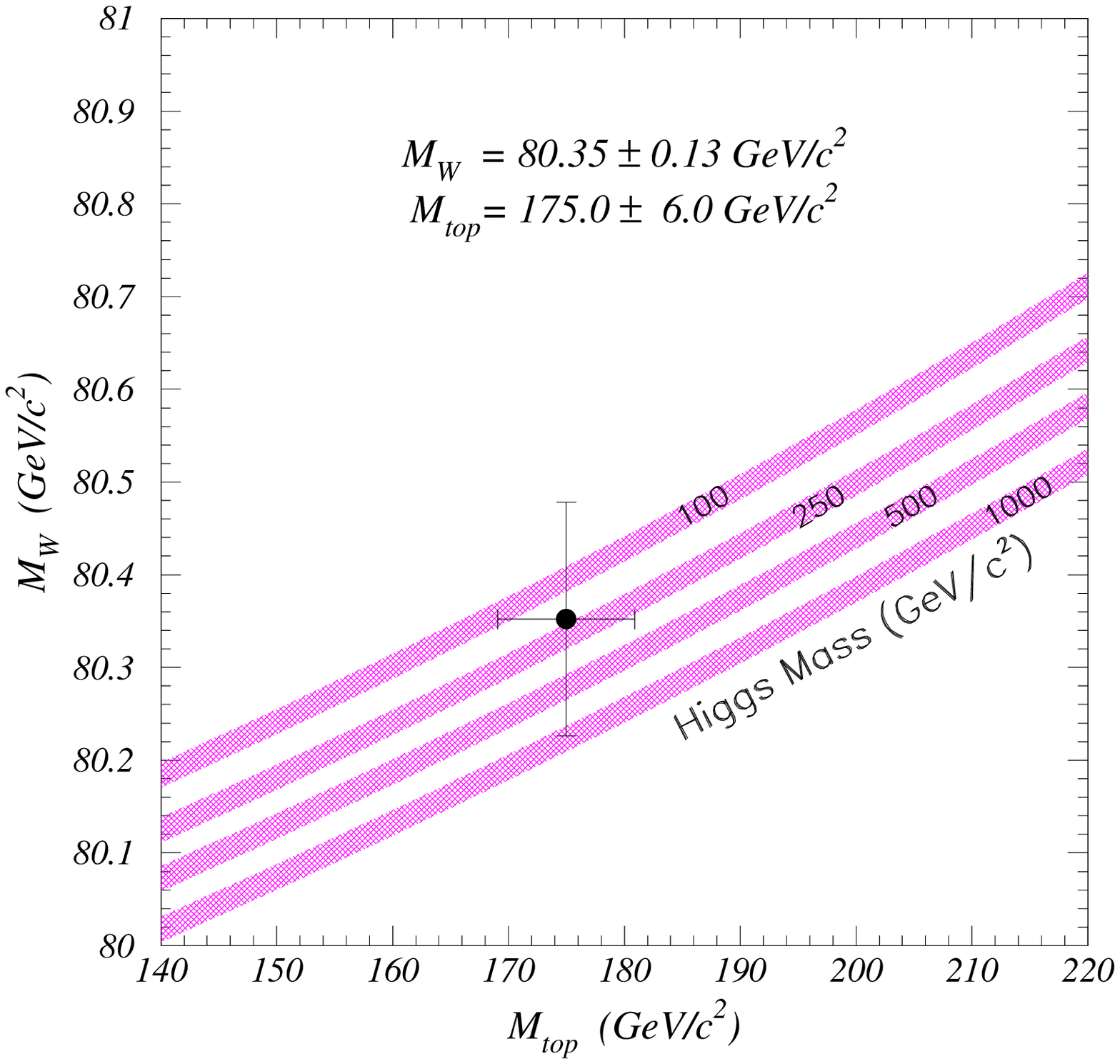}

\smallskip
\parbox{5.5in}{\small Figure 3: The most recent measurements of $m_W$ and $m_{\rm top}$ are confronted with Eq.~(\ref{eq:F(m's)}) for 4 choices of $m_H$ (results presented at the XXVIII International Conference on High Energy Physics, Warsaw, Poland, 1996.)}
\end{figure}

Failure to do so will undoubtedly raise the question of the precision of the computations. State of the art calculations include all dominant 2-loop effects. This should be sufficient to confront Higgs vacuum polarization effects such as (\ref{eqfig:Dh}) with experiment. Some doubts remain about the accuracy of the $e^+e^-$ data in the vicinity of charm thresholds which are used to evaluate the charm quark contribution to the running of $\alpha$; see (\ref{eq:Deltalpha}). The evaluation of the threshold contribution of the $t\bar t$ loops to the same integral is not totally understood. These most likely represent the true limitation of the calculation but neither problem is likely to preclude the indirect measurement of $m_H$. 

From the theoretical point of view, the present victories are bittersweet. The Standard Model is incomplete --- in gauge theory the only natural values for $m_W$ are zero or the Planck mass. The model does not contain the physics that dictates why its actual value is of order 100~GeV. The Standard Model is actually incomplete in more flagrant ways. Unification is, for instance, in terms of an undetermined parameter $\theta_w$; see (\ref{eq:e=gstw}). Its value is determined by physics beyond the Model for which we have no experimental pointers (except perhaps for some hints of non-vanishing neutrino masses). 

\section*{Acknowledgements}
These lectures benefitted from a critical reading by Ted Allen, John Beacom, Chung Kao, and David Rainwater.
This research was supported in part by the U.S.~Department of Energy under Grant No.~DE-FG02-95ER40896 and in part by the University of Wisconsin Research Committee with funds granted by the Wisconsin Alumni Research Foundation.

\newpage
\section*{Appendix A: Calculating a loop}
\setcounter{equation}{0}
\renewcommand{\theequation}{\arabic{equation}A}

How to compute $b$, formally introduced in (\ref{eq:beta-2}), or $\pi(q^2)$ is clear. The answer is
given by (\ref{eqfig:pi(q^2)}), (\ref{eq:alpha-2}) and (\ref{eq:alpha-QED}). The UV cutoff $\Lambda$ removes the infinite part of the
loop which can, in a renormalizable gauge theory, be absorbed in a redefinition of the bare charge. The latter becomes a parameter to be fixed by
experiment. This is standard old-fashioned QED. Nowadays we avoid the explicit
introduction of a cutoff such as $\Lambda$ in (\ref{eq:pi(q^2)}) which spoils the gauge
invariance of the calculation. One instead uses dimensional regularization to
compute $\pi(q^2)$. The basic idea is to carry out loop integrations in a
space with dimensions $n<4$, where they are finite. The result is then
analytically continued to $n=4$ where the UV divergent part of the loop
appears as a $1/(n-4)$ pole. Propagators and interaction vertices remain
unchanged. E.g.\ for the loop (\ref{eqfig:pi(q^2)}) 
\begin{equation}
\begin{array}{r}
\ds -i(g_{\mu\nu}q^2 -q_\mu q_\nu)\pi(q^2)=\left(e_0\mu^{4-n\over2}\right)^2(-1)
\int{d^n k\over(2\pi)^n}\,{\rm Tr} {\gamma_\mu(\ksl+m)\,
\gamma_\nu(\qsl+\ksl+m)\over\left[(q+k)^2-m^2\right]\left[k^2-m^2\right]}\;,\\
\noalign{\vskip2ex}
\ds =-4e_0^2\mu^{n-4}\int_0^1dx\int{d^nQ\over(2\pi)^n}\,
{g_{\mu\nu}\left[{2-n\over n}Q^2+m^2+q^2x(1-x)\right]-2q_\mu q_\nu x(1-x)
\over\left[Q^2-\bigl(m^2+q^2x(x-1)\bigr)\right]^2}\;.  %(2.18)
\end{array} \label{eq:loop}
\end{equation}
Here $m$ is the electron mass and $k$ the 4-momentum circulating in the loop.
The only modification is the introduction of the 't~Hooft mass $\mu$
introduced as a factor $\mu^{(4-n)}$ in order to keep the coupling constant
dimensionless. The last relation follows by executing the following steps:

\noindent
i) use
\begin{equation}
{1\over ab}=\int_0^1dx\,{1\over\bigl[ax+b(1-x)\bigr]^2}\;,
\end{equation} %(2.19)
ii) change variable
\begin{equation}
 Q=k-qx\;, 
\end{equation} %(2.20)
iii) do the traces as usual, but notice that
\begin{eqnarray}
 \gamma_\mu\gamma^\mu&=&n\ 1\;, \\ %(2.21)
 \gamma_\mu\gamma_\alpha\gamma^\mu&=&(2-n)\gamma_\alpha\;. %(2.22)
\end{eqnarray} 
From (\ref{eq:loop}) we then find that
\begin{equation}
\pi(q^2)={8e_0^2\mu^{(4-n)}\over(16\pi^2)^{n\over4}}
 \int_0^1dx\,x(1-x)\left[m^2+q^2x(x-1)\right]^{{n\over2}-2}
\,\Gamma\!\left(2-{n\over2}\right)   \label{eq:piq2}
\end{equation} %(2.23)
by using the relation
\begin{equation}
 \int{d^nQ\over(2\pi)^n}\,{1\over(Q^2-C)^2} = {i\over(16\pi^2)^{n\over4}}
\, \Gamma\!\left(2-{n\over2}\right)C^{\left({n\over2}-2\right)}\;. \end{equation} %(2.24)
We now make a Taylor expansion of (\ref{eq:piq2}) around $n=4$ using the following relations:
\begin{eqnarray}
 \mu^{(4-n)} &=& 1+{4-n\over2}\ln\mu^2+\cdots\ , \\ %(2.25\rm a)\cr \nv6
(16\pi^2)^{n\over4}&=&16\pi^2\left(1+{n-4\over2}\ln4\pi+\cdots\right)\ ,\\
%(2.25\rm b)\cr \nv6
\Gamma\left(2-{n\over2}\right)&=&-{2\over n-4}-\gamma_E^{\vphantom y} \,(=0.5772\ldots)\ ,\\
%(2.25\rm c)\cr\nv6
C^{\left({n\over2}-2\right)}&=& 1+{n-4\over 2}\ln C+\cdots\ . %(2.25\rm d)\cr}
\end{eqnarray}
We thus obtain the desired separation of the $n=4$ infinite and finite parts of
$\pi(q^2)$ with
\begin{equation}
\begin{array}{rl}
\ds \pi(q^2)={\a\over3\pi}\biggl[\!\!\!\! &\ds -{2\over n-4} -\gamma_E^{\vphantom y} +\ln4\pi\\
\noalign{\vskip1ex}
&\ds -6\int_0^1dx\,x(1-x)\ln\left({m^2+q^2x(1-x)\over\mu^2}\right)+\O(n-4)
\biggr]\;,  % \eqno(2.26)
\end{array}  \label{eq:sep}
\end{equation}
which yields (\ref{eq:foo}) in the limit of large $(-q^2)$.

In old-fashioned QED the renormalized charge (Thomson charge at $q^2=0$) would
be defined as
\begin{equation}
 e^2\equiv{e_0^2\mu^{n-4}\over1+\pi(0)}\;,\qquad \alpha\equiv{e^2\over4\pi}
\end{equation}   %(2.27)
with $\pi(0)$ given by (\ref{eq:sep}). In the modern approach, previously introduced, vacuum polarization effects are completely absorbed in the ``running'' renormalized coupling by allowing $\mu$ to vary. $\alpha(\mu)\equiv
e^2(\mu)/4\pi$ is related to $\alpha$ by
\begin{equation} {1\over\alpha(\mu^2)}-{1\over\alpha}=-{1\over3\pi}\ln\left(\mu^2\over m^2
\right)\;.     \label{eq:related}
\end{equation} %(2.28)
Equation~(\ref{eq:related}) implements the so-called $\overline{\rm MS}$ renormalization scheme
where the term ${\gamma_E^{\vphantom y}\over2}-{1\over2}\ln4\pi$ are subtracted out along with
the $1\over n-4$ pole into the renormalized charge. We have now succeeded in
computing $b$ appearing in the formal relation (\ref{eq:beta-2}). Eq.~(\ref{eq:related}) just evolves
the $\overline{\rm MS}$ charge from $Q^2=0$ to $Q^2=\mu^2$ and one sees that
$b={1\over3\pi}$. If $\mu^2$ is such that other loops of leptons and quarks
contribute then
\begin{equation}
 {1\over\alpha(\mu^2)}={1\over\alpha}-{1\over3\pi}\sum_f
Q_f^2\,\ln\!\left(\mu^2 \over m_f^2\right)\;,
\end{equation} %(2.29)$$ 
where the sum is over all fermions with charge $Q_f$. For the traditional computation of a loop we refer to Ref.~\cite{Books}.

\newpage

\section*{Appendix B: Asymmetries at the $Z$-pole}
\setcounter{equation}{0}
\renewcommand{\theequation}{\arabic{equation}B}

Equation~(\ref{eq:ALR}) is valid near $q^2\simeq z$ with 
\begin{eqnarray}
A_{\rm FB}&=&\left( \int_0^1d\cos\theta {d\sigma(\epem\to f\bar f)
\over d\cos\theta} -\int_{-1}^0d\cos\theta{d\sigma(\epem\to f\bar f)
\over d\cos\theta }\right)\Big/\sigma(e^+e^-\to f\bar f)\;,\qquad\\ 
A_{\L\R}&=&\biggl(\sigma\left(\epem_\R\to f\bar f\right)-\sigma\left(\epem_L\to 
f\bar f\right)\biggr)\Big/\sigma\left(e^+e^-\to f\bar f\right)\;,\\
A_\tau&=&\biggl(\sigma\left(\epem\to\tau_\L^-\tau^+\right)-\sigma\left(\epem\to
\tau_\R^-\tau^+\right)\biggr)\Big/\sigma(e^+e^-\to \tau\bar\tau)\;.
\end{eqnarray}
In the above asymmetries $\theta$ is the angle between the produced fermion $f$
and the incoming $e^-\,e^-_{\L,\R}$ represent left- and right-handed
longitudinally polarized electrons and $\tau^-_{\L,\R}$ left- and right-handed
$\tau$'s whose polarization can be experimentally analyzed by observing the
decay $\tau\to \pi\nu_\tau$.


\begin{thebibliography}{9}

\bibitem{Books}
Books nicely complementing this discussion are:\\
 M.~Veltman, ``Diagrammatica", Cambridge Lecture Notes in Physics~4, Cambridge\break University Press;\\
F. Halzen and A. D. Martin, ``Quarks and Leptons", Wiley and Sons, New York;\\
E. Leader and E. Predazzi, ``An Introduction to Gauge Theory", Cambridge Monographs on Particle Physics, Nuclear Physics and Cosmology~3, Cambridge University Press.

\bibitem{ParticleWorld}
For an analytic expression of $\Delta r$ to leading order, see F.~Halzen and D.~Morris, Particle World, Vol.~2, No.~1, pp.~10--20 (1991).

\bibitem{Brazil}
F. Halzen, B. Kniehl, and M. L. Stong, ``The Standard Electroweak Model: Quantum Corrections and Symmetry Breaking", in Particle Physics, {\it Proceedings of the VI~Jorge Andr\'e Swieca Summer School}, edited by O.~J.~P.~Eboli, M.~Gomes, and A.~Santoro, World Scientific (1992).

\end{thebibliography}
\end{document}